\documentclass[useAMS,usenatbib]{mn2e}
\usepackage{graphicx}

\title[Transmission spectrum of HD 189733]{Detection of atmospheric haze on an extrasolar planet: The 0.55 -- 1.05 micron transmission spectrum of HD189733b with the Hubble Space Telescope}
\author[F. Pont et al]{F. Pont$^{1}$, H.  Knutson$^{2}$, R.  L. Gilliland$^{3}$, C. Moutou$^{4}$, D.  Charbonneau$^{2}$\\
$^{1}$Geneva University Observatory, 1290-Sauverny, Switzerland\\
$^{2}$Harvard-Smithsonian Center for Astrophysics, 60 Garden Street, Cambridge, MA 02138, USA\\
$^{3}$Space Telescope Science Institute,3700 San Martin Drive, Baltimore, MD 21218, USA\\
$^{4}$Laboratoire d'Astrophysique de Marseille, Traverse du Siphon, 13376 Marseille Cedex 12, France
}  

\begin{document}

\bibliographystyle{mn2e}



\maketitle

\label{firstpage}

\begin{abstract}

The nearby transiting planet HD~189733b was observed during three transits with the ACS camera of the Hubble Space Telescope in spectroscopic mode. The resulting time series of 675 spectra covers the 550-1050 nm range, with a resolution element of $\sim 8$ nm, at extremely high accuracy (signal-to-noise ratio up to 10,000 in 50 nm intervals in each individual spectrum). Using these data, we disentangle the effects of  limb darkening, measurement systematics, and spots on the surface of the host star, to calculate the wavelength dependence of the effective transit radius to an accuracy of $\sim$ 50 km. This constitutes the ``transmission spectrum'' of the planetary atmosphere. It indicates at each wavelength at what height the planetary atmosphere becomes opaque to the grazing stellar light during the transit.
In this wavelength range, strong features due to sodium, potassium and water are predicted by atmosphere models for a planet like HD 189733b, but they can be hidden by broad absorption from clouds or hazes higher up in the atmosphere. 

We observed an almost featureless transmission spectrum between 550 and 1050 nm, with no indication of the expected sodium or potassium atomic absorption features. Comparison of our results with the transit radius observed in the near and mid-infrared (2-8 $\mu$m), and the slope of the spectrum, suggest the presence of a haze of sub-micron particles in the upper atmosphere of the planet. 
 
\end{abstract}



\section{Introduction}

Transiting planets offer a unique opportunity to study the atmospheres of extrasolar gas giants. During the transit, the height at which  the planetary atmosphere becomes opaque to the grazing star light varies with wavelength. Thus the transmission spectrum of the planet's atmosphere along its limb can be recovered from the wavelength dependence of the radius of the planet indicated by the depth of the transit. 
The transit depth change is very small -- typically less than one part per thousand of the stellar flux -- but it is accessible with space telescopes for planets with bright enough host stars. 
Transmission spectroscopy provides information on the composition and physical properties of the planetary atmosphere close to the limb. Several theoretical
studies have predicted transmitted spectra of hot Jupiters \citep[e.g.][]{sea00,bro01,bar07,tin07}. In the 550--1050 nm wavelength range covered by the present study, the models predict strong atomic lines at 589 nm from sodium and 769 nm from potassium, and several molecular bands due to water. Further into the red, the model spectra are dominated by molecular bands of water, methane, ammonia and carbon monoxide. The largest degree of uncertainty in the models is believed to be governed by the possible presence of condensates forming hazes or clouds, that can block the stellar flux at any height in the atmosphere. Hazes and clouds produce featureless continuum absorption in the spectrum, with a shape depending on the size distribution of the particles. In close-in gas giants ($T_{\rm eq}=1000-1500$ K), silicates and iron compounds can form solid particles \citep{sea00,for05}. Condensates can also arise from the formation of compounds by non-equilibrium photochemistry in the strongly irradiated upper atmosphere \citep{for03}. 
Strong signatures of clouds and hazes are observed in the spectra of brown dwarfs with similar atmospheric temperatures \citep{hel06b}. 

All planets of the Solar System with a thick atmosphere have spectral properties strongly affected by hazes and clouds. Venus, Titan and Triton have opaque atmospheres in the visible, that become transparent to much deeper layers in the mid-infrared, where the wavelength exceeds the typical size of the condensate particles by a sufficient factor. 

The atmospheric properties of transiting extrasolar giant planets 
are increasingly constrained by new observations. HD~209458 \citep{cha00,hen00} and HD~189733 \citep{bou05} are the most scrutinized
systems because their bright host star ($V\sim 7.7$ mag) and large radius allow precise
enough spectroscopic measurements.
The first positive
application of transmission spectroscopy was the detection of
the NaI absorption signature by the Hubble Space Telescope (HST)
in the atmosphere of HD~209458b 
\citep{cha02}.  This signature is much weaker than expected, which was interpreted as due to the presence of clouds  \citep{for03}, photoionization \citep{bar07} or condensation of sodium on the night side \citep{iro05}. 
So far all ground-based attempts at detecting absorption features have
derived only upper limits \citep[e.g.][]{mou01,bro02,mou03,win04,nar05,dem05,barn07}. The varying spectral signature of the Earth's atmosphere can easily mask the tiny effect of the differential transit radius of the planet. The much higher
precision of data obtained from space is critical in this
search. 

HD~189733$b$ is a gas giant planet orbiting a K dwarf star in 2.2 days \citep{bou05}.  Because the primary star is small (0.755 R$_\odot$), this system offers the advantage of a very deep transit signal (2.4 \% flux drop).
Data obtained by the Spitzer Space Telescope in the mid-infrared on this system allowed the
analysis of the primary transit \citep{knu07,tin07,ehr07}, the secondary eclipse \citep{dem05,gri07,knu07,swa07}, and the day-to-night flux variation
along the orbit \citep{knu07}. 
\citet{tin07} present a complete model of the transmission spectrum fitted to the mid-infrared data (Spitzer bands at 3.6, 5.8 and 8 $\mu$m), and predict very strong features  at shorter wavelengths for an atmosphere without condensates.
An HST lightcurve of HD~189733 in the 550-1050 nm range has been obtained over three transits with the Advanced Camera for Surveys (ACS) and analysed in white light by \citet[][hereafter ``Paper~I'']{pon07}. This data set provides the opportunity to test the prediction of Tinetti et al. in the 600-1000 nm range, and in particular to explore the presence of water absorption and the transparency of the atmosphere of the planet. The signal-to-noise ratio of the ACS spectra is much higher than the Spizer mid-infrared lightcurves because of the larger size of HST and the higher stellar flux in the visible.
The ACS data is precise enough to distinguish the signature of a clear atmosphere with water and alkali metal absorption features from that of clouds or hazes blocking the stellar flux higher up in the atmosphere.

In this paper, we present the spectroscopic analysis of the ACS/HST data set, in
order to constrain the transmission spectrum of the planetary atmosphere from 550 to 1050~nm.

\section{The data set }

\begin{figure}
\resizebox{8cm}{!}{\includegraphics{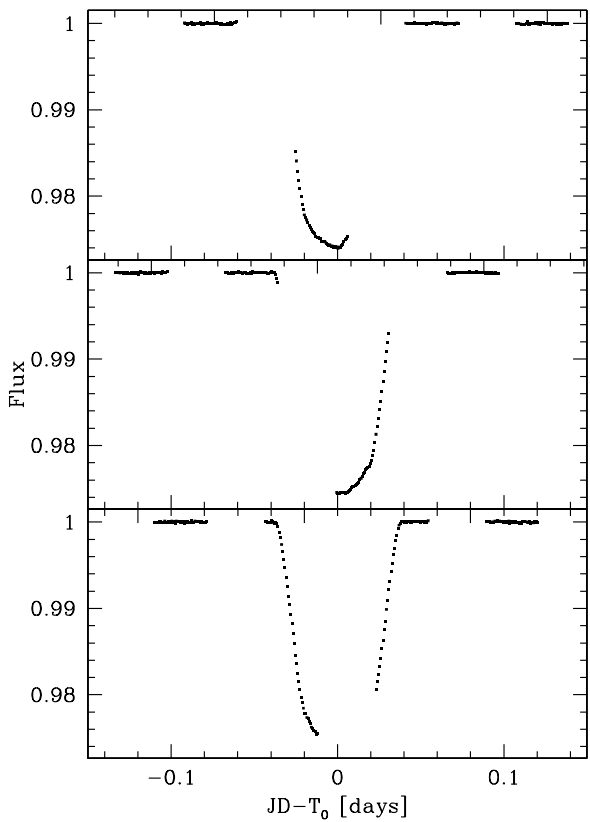}}
\caption{The HST/ACS lightcurve integrated over all wavelengths. In the middle of the first transit, the planet occults a large starspot, producing a steep rise in intensity at the end of the second HST orbit. Smaller structures on the star are also crossed during the second transit. See Paper~1 for a detailed analysis.}
\label{figlc}
\end{figure}

The data set analysed in this paper consists of three time series of ACS low-resolution spectra  covering five HST orbits each, during three transits of the HD 189733 system. A total of 675 spectra were measured, covering the 550-1050 nm wavelength range, with a resolution element of about 8 nm. The accuracy on the relative intensity  integrated over all wavelengths is $7 \times 10^{-5}$ on individual exposures, and $3 \times 10^{-5}$ on 10-minute averages, including correlated non-photon sources of noise.
 The data is presented in detail in Paper~I, together with the analysis of the lightcurve obtained by integrating the first-order  ACS spectra over all wavelengths, and decorrelating for instrumental systematics related to changes in the telescope and detector conditions. 
 Figure~\ref{figlc} shows this ``white'' lightcurve (i.e. integrated over all wavelengths) of our ACS/HST data. The only significant departures from a transit curve model are observed near the middle of the transit, and attributed to the occultation of starspots by the planet. In the present paper, we analyse the chromatic information in the spectroscopic time series, with the objective of measuring the transmission spectrum of the planetary atmosphere during the transits. We use the 50-nm-wide binning of the spectrum described in Paper~I, i.e. our primary data set consists of ten time series, each covering a 50-nm wavelength interval from 550 nm to 1050 nm. These are obtained by restricting the extraction to specific column ranges. We also repeated the reduction using the finest sampling possible, 1-pixel wide intervals, which corresponds to about 3-nm bands in wavelength.  This produces 173 passbands from 537 to 1069 nm. The chromatic data is decorrelated from instrumental systematics in the same manner as the white lightcurve. The fine-grain data set is useful to examine the structure of narrow spectral features, namely the sodium and potassium atomic lines. The coarser-grain extraction (50-nm bands) is more robust against errors in the decorrelation of systematics, and we use it to measure the overall shape of the spectrum and the effect of molecular bands.

\section{Data Analysis}
\label{ana}
\subsection{Overall strategy}

The depth of the dimming due to a planetary transit is directly proportional to the projected surface of the planet relative to the star, which grows with the square of the planetary radius.
In a monochromatic lightcurve, the transit depth measures the square of the radius at which the planetary atmosphere becomes opaque to the grazing star light for a specific wavelength. The transmission spectrum of the planet is therefore obtained by fitting the same transit curve model for different wavelengths, leaving the planetary radius as a free parameter.  In wavelengths where the absorption of stellar photons passing through the planetary atmosphere is more efficient, the planet appears to have a larger radius. 
The principle is simple enough, but requires data of exquisite quality, because the signal represents only a tiny fraction of the signal of the transit (typically $10^{-2}$ to $10^{-3}$), which is itself a small modulation of the star's lightcurve. 

Among known transiting planetary systems, HD~189733 is the system where the transmission spectrum is expected to be most accessible to observations among known transiting planets, because the radius ratio is large, the planet has a hot H$_2$ atmosphere with a large scale height, and the star is bright. Signatures of radius change in the transmission spectrum are expected to amount to $10^{-3} - 10^{-4}$ of the stellar intensity.
However, the case of HD 189733 is complicated by the fact that the host star is variable. It is active at the few percent level in visible photometry, with obvious modulations of the lightcurve due to rapidly-varying starspots, rotating with the star in about 12 days \citep{cro07,hen07}. The presence of spots has two effects on the measurement of the transmission spectrum: (1) the occultation of spots by the planet during the transits introduces features in the transit lightcurve (2) unocculted spots on the surface of the star modify the relation between the transit depth and the radius ratio. Both effects are wavelength-dependent, and must therefore be taken into account not only as a global correction to the transmission spectrum, but as a passband-to-passband correction.

The challenge is to disentangle properly the four following causes for the variability of the spectroscopic time series:
\\
-- Main transit parameters : period, epoch, system scale ($a/R$), orbital inclination angle\\
-- Telescope/detector systematics\\
-- Limb darkening\\
--  Spots and variability\\
--Wavelength dependence of the radius ratio (transmission spectrum)\\

Thanks to the extremely high signal-to-noise ratio of the ACS data, a sufficient precision on the transmission spectrum can be achieved to detect the main spectral features predicted by planetary atmosphere models. 
The signal-to-noise ratio on individual data points in each 50-nm passband is in the 4000:1 -- 10000:1 range. With 675 images and 10 passbands, there is an enormous amount of information in the data itself.

In the following four subsections we consider the causes of variability one by one, first explaining how we tackle them, and then estimating what level of systematic uncertainties they are likely to leave in the measurement of the transmission spectrum of HD 189733b.

\subsection{Main transit parameters}

The orbital period, phase and inclination angle, and the system scale, do not depend on the wavelength. We fix the values of these parameters to those derived in Paper~I from the lightcurve integrated over all wavelengths (``white'' lightcurve).

\begin{figure*}
\resizebox{16cm}{!}{\includegraphics{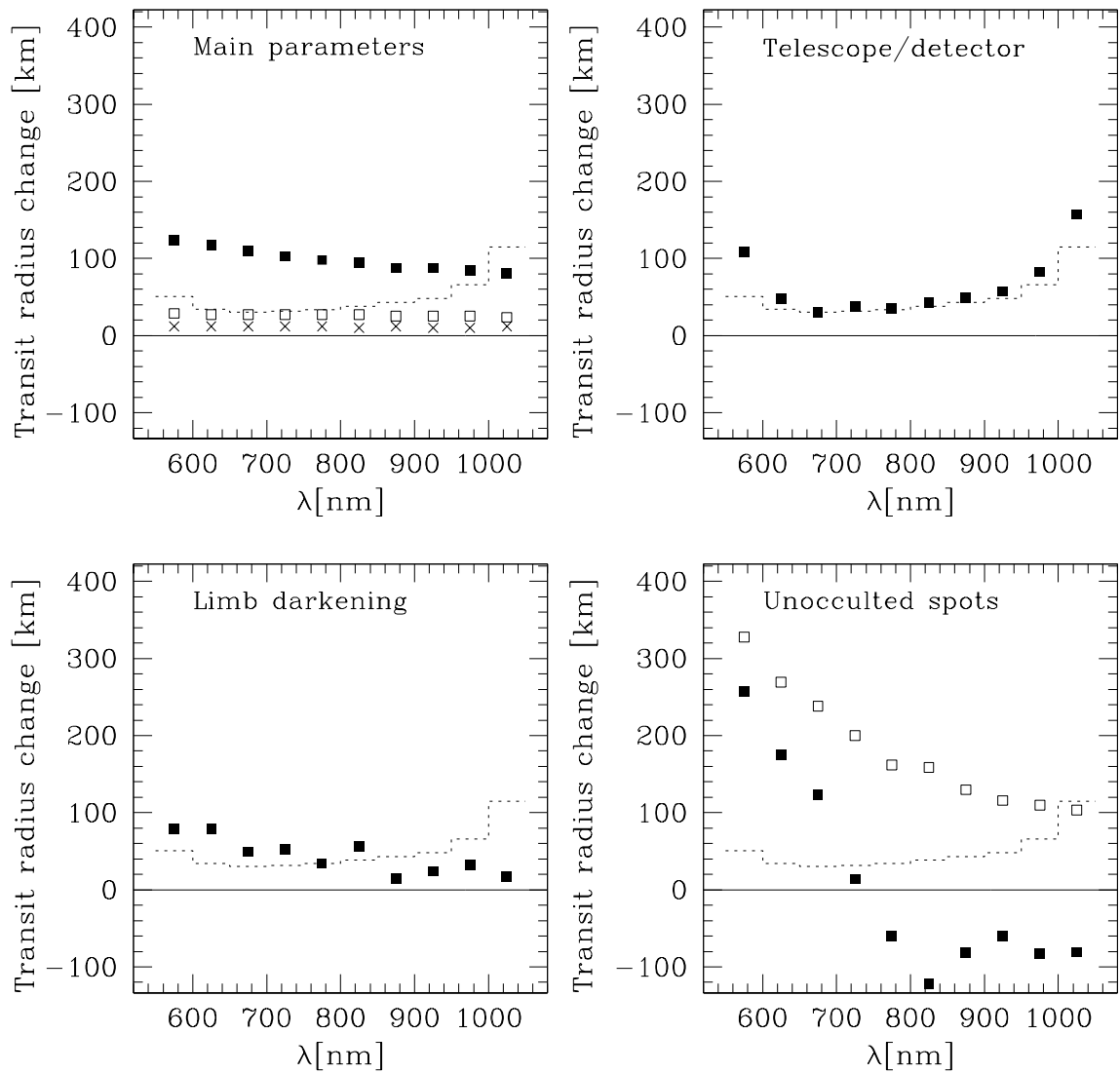}}
\caption{Changes in the transmission spectrum results caused by the uncertainties in the different steps of the analysis. The vertical scale shows the change in planet transit radius. The dotted bins in all panels show the photon shot noise. {\bf Top left:} effect of moving the transit timing (crosses), radius ratio (open squares) and inclination angle (solid squares) by the 1-$\sigma$ uncertainty interval in Paper~1. {\bf Top right:} effect of residual red noise left over from the decorrelation of instrument/detector parameters, using the method of \citet{pon06}. {\bf Bottom left}: effect of shifting the limb darkening coefficients by one passband. {\bf Bottom right}: effect of adding unocculted spots absorbing 1\% of the stellar flux (open squares), and of changing the mean temperature of the spots from 4000 K to 3500 K (solid squares).}
\label{figsyst}
\end{figure*}

The effect of the uncertainties on these parameters can be evaluated by moving them by their $\pm 1$-$\sigma$ error interval and repeating the whole analysis with these new values. The result of this procedure is shown on the upper left panel of Fig.~\ref{figsyst}. Changing the transit epoch and period has little effect on the transmission spectrum. Modifying the inclination angle has more impact, both on the absolute value and on the slope, because the run of limb darkening along the chord of the projected path of the planet on the star changes.
In all cases, the uncertainties on the main transit parameters produce changes with a very smooth wavelength dependence,  so that they cannot introduce or mask narrow features in the transmission spectrum.

\subsection{Telescope and detector systematics }

Systematic flux variations in the ACS data were removed with decorrelation using six vectors: the $x$ and $y$ position on the chip, the spectrum width, angle, the HST orbital phase, and linear time. The details are given in Paper~I. The passband-integrated intensities were decorrelated in the same way as the white lightcurve, on each passband independently. We did not attempt to take into account the covariance between the corrections of neighbouring passbands. 

Residual systematics can be estimated from the level of correlated noise in the extensive part of the time series that was taken outside the transit. 
The top right panel of Fig.~\ref{figsyst} shows the amplitude of the residual systematics in the ten passbands that we use, using the method of \citet{pon06} to estimate the effect of red noise on the planetary parameters. They have an amplitude comparable to the photon noise, except at the edges. The shortest and longest passbands are more sensitive to systematics because the actual flux on the detector drops sharply on the edges of the spectrum, which amplifies the effect of telescope pointing drifts and focus changes. 

Three different transits are covered by our data, so that the instrumental systematics can also be estimated from the differences between the transmission spectrum measured in the three visits. The systematics are not correlated between the visits, because they were taken at three very different positions in terms of HST configuration. Of course, this estimate will be an upper limit, because other factors affect the comparison between the three transits (namely the starspots, see below), and the second and third visits have fewer points during the transit. The comparison between the results from the three visits is discussed in Section~\ref{results}.

\subsection{Limb darkening}
\label{limbd}

Limb darkening has an important effect on the transit curve shape in the wavelength range covered by the data, and is strongly coupled with the transit depth. Fortunately, limb-darkening is a well-understood effect and can be modelled with high accuracy for a K dwarf in this range of wavelengths. We used a synthetic stellar spectrum calculated by R. Kurucz for $T_{eff}=5050$ K, $\log g=4.53$ and [Fe/H]=$-$0.03 \citep[from][]{bou05}\footnote{The synthetic spectrum for HD 189733 is available on the website of R. Kurucz, http://kurucz.harvard.edu/stars/HD189733.}, to compute the intensity in small wavelength intervals and at 17 radial positions spanning the region from the center out to the edge of the star. We then calculated the flux-weighted average intensity distribution across the surface of the star in each of our passbands, and fit this intensity distribution with the four-parameter non-linear limb-darkening law given in \citet{cla00} ($I(\mu)=1-\sum^{4}_{n=1} c_n (1-\mu^{n/2})$) to determine the limb-darkening coefficients for each passband. The corresponding limb-darkening coefficients are listed in Table~\ref{ldk}.

\begin{table}
\centering
\begin{tabular}{c c c c c}
Passband [nm] &    c1   &        c2  &        c3   &       c4   \\ \hline \hline
550 -- 600 & 0.4621    &$-$0.2003   & 0.9450    &$-$0.4045 \\
600 -- 650 & 0.5148   & $-$0.2774 &   0.9429   & $-$0.4033 \\
650 -- 700 &  0.5620   & $-$0.3531  &  0.9524  &  $-$0.4096 \\
700 -- 750 & 0.5932   & $-$0.4183  &  0.9488   & $-$0.4003 \\
750 -- 800 & 0.6231   & $-$0.4808  &  0.9672   & $-$0.4078 \\
800 -- 850 & 0.6390   & $-$0.5260  &  0.9695   & $-$0.4039 \\
850 -- 900 & 0.6529   & $-$0.5366  &  0.9290   & $-$0.3886 \\
900 -- 950 & 0.6538   & $-$0.5508  &  0.9273   & $-$0.3867 \\
950 -- 1000 & 0.6475   & $-$0.5439  &  0.9104   & $-$0.3837 \\
1000 -- 1050 &  0.6303  &  $-$0.4967 &   0.8538  &  $-$0.3696 \\ \hline
\end{tabular}
\caption{Flux-weighted limb darkening parameters for our passbands calculated from a Kurucz model atmosphere.}
\label{ldk}
\end{table}

To estimate the potential systematic uncertainties introduced on the final spectra by the choice of limb-darkening coefficients, we computed the spectrum using  for each passband the limb-darkening coefficients appropriate for the next passband. The bottom left panel of Fig.~\ref{figsyst} shows the difference between the resulting spectrum and the one using the limb-darkening parameters of Table~\ref{ldk}. The effect of this rather large simulated uncertainty on the limb-darkening parameters is to shift the whole spectrum by an amount comparable with the photon noise. As expected, a change in limb-darkening parameters affects the overall slope of the spectrum, via the covariance between limb darkening and transit depth. Spectral features, however, are not affected, because the intensity and shape of limb darkening varies rather smoothly in the $\lambda=600-1000$~nm range at that spectral resolution. 


\subsection{Starspots}
\label{spots}

The presence of starspots on the surface of HD 189733 affects the measurement of the planetary transmission spectrum in two ways: 

\noindent
(a)- spots occulted by the planet modify the shape of the transit lightcurve. The modification is chromatic, since the occulted spots have a different spectrum from the rest of the star

\noindent
(b)- spots {\it not} occulted by the planet, but visible to the observer, diminish the brightness of the unocculted part of the star compared to a spot-free surface, and therefore slightly increase the depth of the transit. This is also a chromatic effect.

One big starspot and a series of small spots are occulted during our time series, and extensive ground-based and space-based follow-up constrains the distribution of spots on the whole surface during the HST observations \citep[][see Paper~I]{cro07,hen07}. There is therefore enough information available to correct for the effect of the spots.

\ \\
\noindent
{\it (a) Spots occulted by the planet}

The planet occulting a spot on the surface of the star during the transit produces a rise in the lightcurve, because the spot is cooler and fainter than the rest of the star.  Hiding the spot therefore increases the overall luminosity of the part of the star not hidden by the planet. This effect varies with wavelength, according to the difference in surface brightness between the spot and the rest of the stellar surface, i.e. the effective temperature of the spot. To first order, the effect behaves like the difference of the spectra of two stellar photospheres at different temperatures.

In  our reference solution, we simply do not use the part of the time series visibly affected by starspots. This in fact includes more that half of the in-transit data (see Paper I), but there is a sufficient stretch of data in the first HST visit without apparent spot effects to cover all phases of the first half of the transit. Since the transit model is symmetrical around the transit mid-time, this sequence is sufficient to constrain the chromatic dependence of the transit depth (i.e. the transmission spectrum of the planet).

\begin{figure}
\resizebox{8cm}{!}{\includegraphics{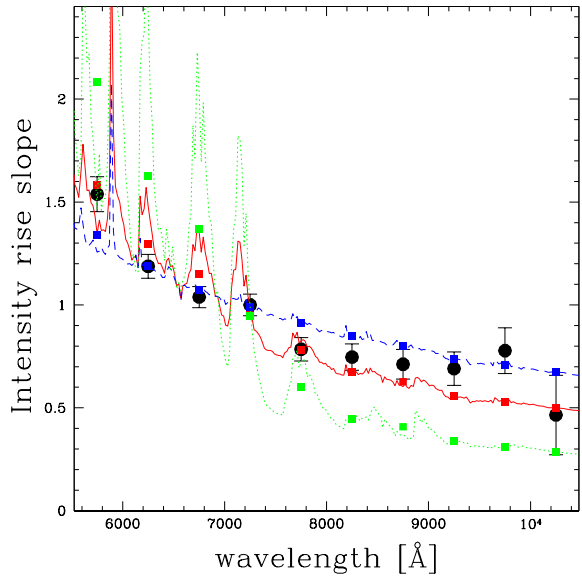}}
\caption{Slope of the relative intensity rise caused by the planet occulting a starspot during the first transit covered by our data, as a function of the central wavelength of the 50-nm spectral bins. The lines show the signals expected for spots with temperatures of 4500 K (dashed), 4000 K (solid) and 3500 K (dotted), with square symbols marking the integration over our passbands, for each model.}
\label{figspots}
\end{figure}

In order to check the robustness of the reference solution, we use the parts of the time series affected by spots to compute independent estimates of the transmission spectrum. To first order, the occultation of the large spot - still ongoing at the end of the HST sequence - corresponds to a linear flux rise. We therefore fit a transit model with a chromatic linear rise to the time series. The free parameters are the transit depth in each passband,  the time of contact with the spot, and the slope of the rise in each passband. The results in terms of slope vs. passband for the spot are shown by the circles in Fig.~\ref{figspots}. 

This procedure has an important side benefit: it provides the spectral dependence of the effect of the occultation of the spot. We can compare the results with the effect expected from occulting a stellar surface of lower temperature against the background of a $T=5000$ K stellar photosphere. The expectations for synthetic Kurucz stellar atmosphere models are shown in Fig.~\ref{figspots}, for spot temperatures of 4500~K, 4000~K and 3500~K, assuming T$_{\rm eff}=5000$~K for the star as a whole. The intensity rise slope is not a smooth function of wavelength, because of the different strength of temperature-dependent absorption features on the photosphere and in the spot. From Fig.~\ref{figspots}, we can estimate the effective temperature within the spot. The observed chromatic effect of the spot corresponds to the expectations from the occultation of a cooler surface with effective temperature between 4500~K and 4000~K . A lower spot temperature of 3500~K predicts a stronger difference than observed in the blue, due to the quasi-disapearance of the blue part of the spectrum under molecular bands at these temperatures.  This range of temperature is consistent with what is known for spots on solar-type stars: large starspots are 500-1000~K cooler than the photosphere in active stars \citep{fra04}.

The effect of smaller starspots occulted during the second HST visit is too small to be modelled in this way. We can use the model of the large spot, however, to correct the effect of the small spots differentially. We assume that the small spots have a similar temperature as the large one. We compute, on the ``white'' lightcurve, the difference between the time series affected by spots, and the theoretical transit curve with the parameters of Paper I. We then remove this difference from the coloured lightcurve, multiplied by the passband dependence obtained for the large starspot. Namely, if $f_i$ is the slope of the rise due to the large spot in passband $i$, $f$ the slope in the white lightcurve, the vector $\Delta F=F_o-F_c$ is the flux difference between the observed white lightcurve and the transit model, then the corrected lightcurve for passband $i$ is:
\[
F'_i = F_i - \Delta F \times f_i/f
\] 
Obviously, the assumption that the spots have identical temperature can be questioned. 
In our case $\Delta F$ is of the order of $5\cdot 10^{-4}$ at most, and $f_i/f$ varies between 0.8 and 1.2. Consequently the uncertainty on the exact spot temperature is a small effect ($<< 2 \cdot 10^{-4}$).

\ \\
\noindent
{\it (b) Spots not occulted by the planet}

Even spots not occulted by the planet affect the transit signal.
Spots make the unocculted part of the star dimmer than a spot-free surface, and therefore increase the depth of the transit and the calculated radius ratio. This also depends on wavelength, since the dimming due to spots depends on the difference of their spectrum with that of the rest of the stellar surface. Such effects are very tiny ($<10^{-4}$ on the wavelength dependence of the transit depth even for an active star like HD~189733), but they need to be taken into account at the level of accuracy reached by our data.

\begin{figure*}
\resizebox{16cm}{!}{\includegraphics{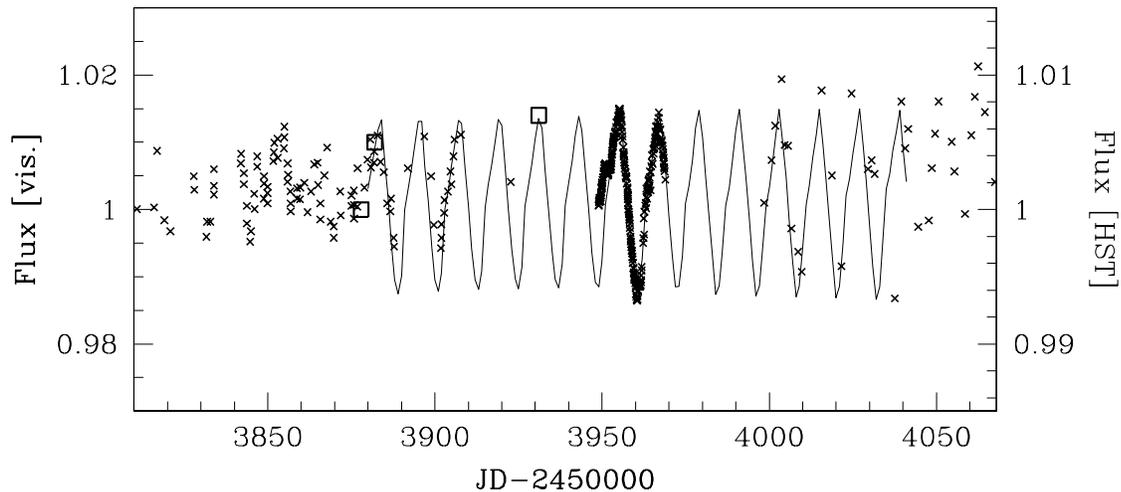}}
\caption{Lightcurve of HD189733 over 260 days. {\it Crosses:} Ground-based monitoring from \citet{hen07} and continuous monitoring with MOST \citep{cro07}. {\it Squares:} Absolute flux  during the three HST visits from Paper~I. The zero-points of the three lightcurves have been adjusted. The vertical scale is doubled for the HST data, to account for the wavelength dependence of spot-related variability. The line shows an extrapolation of the MOST lightcurve over 6 cycles before and after the actual measurements.}
\label{figphot}
\end{figure*}

In Paper~I, we accounted for this effect differentially, using the variations of the total flux due to spot extinction between the three transits covered by the data. The flux variations are measured by the absolute calibration of the HST data itself, and by the simultaneous ground-based monitoring of the star. Figure~\ref{figphot} places the three HST measurements in the context of the long-term monitoring by \citet{hen07} and the observing campaign of the MOST space telescope \citep{cro07}. The ground-based and MOST data are taken at similar wavelengths, centered near 560 nm. To account for the much redder position of the HST spectra, the intensity scale was expanded by a factor 2. This factor was found by modelling the spectral distribution of the effect of spots at 4000 K on a stellar surface at 5000 K.
In the figure, one complete cycle of the MOST data is repeated six times before and after the actual coverage. This extrapolation connects smoothly with the ground-based data of the previous and subsequent season both in amplitude and phase, which suggests that the pattern of spots has remained relatively stable. In contrast, just before our first observations, the ground-based photometry indicates a change of behaviour of the spot pattern in less than one rotation cycle.
The three absolute flux measurements from the HST data, showing variation of $+5\times 10^{-3}$ in flux between the first and second visit, and  $+7\times 10^{-3}$ between the first and third visit (Paper~I), are in agreement with this picture. 

For the chromatic data, we need not only an estimation of the flux difference caused by star spots moving in and out of view during the visits, but also of the total flux dimming caused by all unocculted spots visible on the stellar surface at the time of the transit, because these spots diminish the flux of the star throughout the transit, and therefore the transit radius calculated from the lightcurve is artificially increased. To account for this effect, we have to estimate the ``background level'' of starspots, i.e. the minimum quantity of dimming due to spots whatever the viewing angle. 


Croll et al. (in prep.) present a spot model for the MOST lightcurve covering two rotation cycles of HD189733 a short time after our sequence, and find that a good fit can be produced by a two-spot model, with a negligible background spot value. The third HST visit is near the epoch when the two large spots are almost entirely hidden.
However, as pointed out by these authors, more complex models with numerous, smaller spots, and time variations in spot temperature and size, or differential rotation velocities for spots as a function of the latitude on the star, could also account for the observed lightcurve. In particular, a higher angle between the stellar spin and the orbital axis of the planet is required in the Croll et al. model than observed by \citet{win07} using the Rossiter-Mc Laughlin effect. A model with more than two spots would be needed to reconcile the two pieces of information. Our ACS time series tends to indicate that smaller spots are indeed present on other parts of the star.

We have constructed several suites of spot models fitting the MOST lightcurve, and have found the following statistical indications: if a few large spots dominate the signal, then a background spot level between 0 and 1 \% is typical (i.e. the spots diminish the intensity by 1\% in all viewing angles compared to a spot-free surface). For models with a large number of spots following a power-law size distribution with -1 slope, the typical background spot level is 1-3 \%.

We can also try to estimate the background spot level from the HST data itself. Let us consider  that the whole surface of the star is covered with small and medium spots like the ones crossed during the first two HST visits. The occultation of these spots causes an average decrease of about $3\times 10^{-4}$ in the transit lightcurve, or $1.2 \% $ of the flux occulted by the planet. Extending this to the whole stellar surface directly translates into a $1.2\%$ factor for the dimming due to background spots if none of the observed photometric variation is due to the distribution of these spots, but entirely to bigger spots. 

\begin{figure}
\resizebox{8cm}{!}{\includegraphics{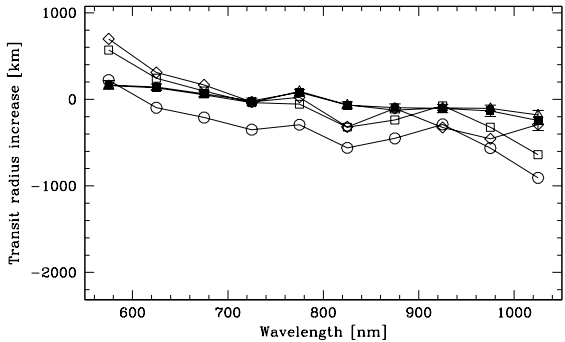}}
\caption{Planetary transmission spectra obtained from four different solutions: first orbit without spot (reference solution `A'; {\it solid squares}), second orbit with empirical correction of occulted spots ('C'; {\it diamonds}), third orbit ('D'; {\it open squares}), and second orbit including raw occulted spots signal ('E'; {\it circles}). Solution `B' is nearly identical to solution A and is not shown.}
\label{figsol}
\end{figure}

There is a shortcut to control the potential effect of unocculted starspots in this scenario. If we assume that the zone of the surface of the star crossed during the second visit is typical of the whole stellar surface, then we can fit for the effective transit radius ratio without any correction for spots. The mean effective radius ratio during this photometric sequence will be an unbiased estimator of the real radius ratio, since the flux dimming in the part of the star occulted by the planet (including the spots) will correspond to the dimming by spots of the unocculted part. This will be valid in all passbands, provided all the spots have similar temperatures. The resulting transmission spectrum will of course be noisier, because the spots affect the accuracy of the description of the lightcurve by a smooth transit model, but the overall level will be correct. The result of using this procedure is shown in Fig.~\ref{figsol} (solution `E'). Note that this ``worst case scenario'' is physically unexpected. Spots in active stars are expected to be large and concentrated at certain latitudes, not covering the entire surface \citep{gra00}. 

The effect of the uncertainty on the spot temperature and on the total amount of unocculted spots is illustrated in the bottom right panel of Fig.~\ref{figsyst}. The black symbols show the difference in the transmission spectrum caused by changing the mean temperature of the unocculted spots from 4000 K to 3500 K. The white symbols show the effect of adding a constant background of unocculted spots with a spectral distribution described by the dots in Fig~\ref{figspots}, amounting to a dimming of 1\% over the whole star. 

Figures~\ref{figsyst} and \ref{figsol} show that the correction for unocculted starspots is the dominant source of overall uncertainty on the exact shape of the transmission spectrum. It can affect its level and global slope at the $\sim 400$ km level. A change in spot temperature has a non-linear effect on the shape of the transit spectrum because of the increasing strength of molecular bands below 4000~K.

The consideration of the ACS/HST spectrum in 1-pixel bins ($\sim 3$ nm) allows us to place upper limits on the total level of background starspots and their temperature. The presence of cool starspots on the star introduce spectral features in the apparent radius ratio, because of the temperature dependence of the sodium line at 589 nm, and the presence of molecular absorption bands below 4000 K. The absence of any such features in our observed spectra -- barring a coincidence between the planetary and stellar spectra -- indicates that the background intensity diminution is not higher than about 3 \% for spot temperatures near 4000 K, and 1 \% for temperatures near 3500 K.

Thus, several independent approaches indicate that the background spot level is between zero and 3 \%, with lower values being more likely, and with typical spot temperatures of 4000 K or higher. In our reference solution, we shall assume a value of 1 \% for the unocculted spot level during visit 1.

\subsection{Wavelength dependence of the radius ratio}

We finally compute the planetary transmission spectrum by fitting the ACS time series for the radius ratio, for each colour passband. The transit model parameters are fixed to those obtained in Paper~I from the combined white lightcurve, and the limb darkening coefficients to the values of Table~\ref{ldk}. We use the \citet{man00} theoretical transit shapes. The computed uncertainties take residual systematics noise into account following the prescriptions of \citet{pon06}.  The wavelength-dependent correction for unocculted spots is applied a posteriori, passband by passband. The spot corrections are taken from the empirical method of Section~\ref{spots}, with a 1\%  dimming from unocculted spots. 

We performed five different reductions, labelled A to E. In the different solutions we change the HST visits used, the way the occulted spots are treated, and the method of determination of the limb darkening.  The results are displayed on Fig.~\ref{figsol}, in terms of height of the effective transit radius as a function of wavelength, relative to an arbitrary fiducial level. 

In solution 'A', we use the data from the first HST visit, outside of the feature caused by crossing the large starspot. This series samples an apparently spot-free region, up to the center of the star. It is therefore very sensitive to the effective radius ratio. 

The four other solutions are of lower accuracy since they use shorter or noisier parts of the time series, but they are a very important way to check the overall level of the sum of all the sources of systematics, since they are partly or entirely independent of each-other. 

In solution `B', we use the data in Visit 1 including the occulted spot. The effect of the spot is assumed to be a linear flux increase, with a wavelength dependence constructed from the difference between two synthetic stellar spectra from the Kurucz library, with $T_{\rm eff}=5000$ K for the star and $T_{\rm eff}=4000$ K for the spot. The limb-darkening coefficients are not fixed to the values of Table~\ref{ldk}, but empirically determined from the best fit to the data of the three HST visits, using a quadratic law.

In solution 'C', we use the data in Visit 2. The small fluctuations due to spots are compensated by assuming the same wavelength dependence as the large spot signal observed in Visit 1. The limb-darkening coefficients are taken from  Table~\ref{ldk}. 

In solution 'D', we use the data in Visit 3, with limb-darkening coefficients from  Table~\ref{ldk}. No correction for occulted spots is applied.

Finally, in solution 'E' we use the data in Visit 2 without spot correction (the shortcut explained in Section~\ref{spots}), with limb-darkening coefficients from  Table~\ref{ldk}.

Overall, the difference between the solutions is consistent with the level of systematics estimated from the out-of-transit fluctuations. Solutions B to E are noisier than solution A, as expected from the effects of spots (B, C and E) or the presence of less in-transit data (D). Nevertheless, the main characteristics of the resulting spectrum are observed in all solutions: the absence of strong feature, a negative slope, a rise below 600 nm and near 800 nm, and a decrease beyond 1000 nm. Note that solution B and C are entirely independent (first vs second HST visit, atmosphere models vs fitted limb-darkening parameters,  empirical vs model spot correction). Solution E confirms that the background level of spots should not affect the results fundamentally. Even if the stellar surface were entirely covered with small spots, the effective radius would decrease by no more than 500~km and the global shape would be conserved.

Solution A is chosen as our 'reference solution', because the other solutions are more sensitive to the uncertainties on the properties of the spots (solution B, C and E), or to the values of limb darkening (solution D). Solution A samples all the phases of the transit on a spot-free stellar surface. We choose not to combine solutions from the three visits, since this would introduce an additional dependence on the changing configuration of the unocculted spots. 


\section{Results: Transmission spectrum of the planetary atmosphere}

\label{results}

\begin{figure}
\resizebox{8cm}{!}{\includegraphics{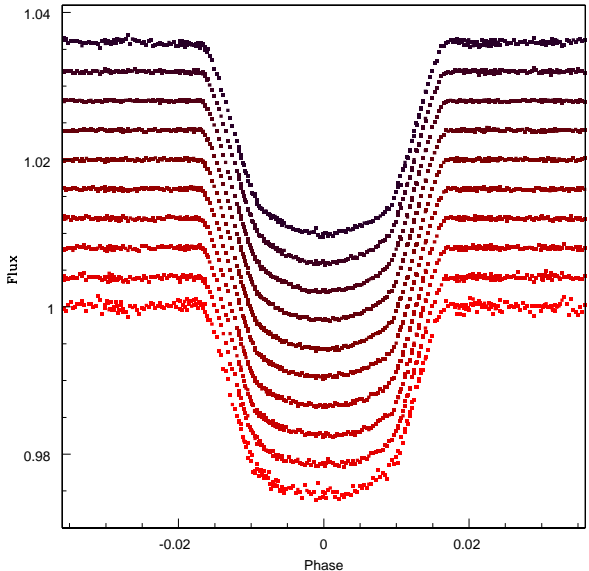}}
\caption{Phased lightcurves in the ten 50-nm chromatic passband from 550 nm to 1050 nm, shifted vertically for visibility. The lightcurves are corrected for the effect of starspots, occulted and unocculted (see text). Longer wavelengths towards the bottom of the figure.}
\label{figchrom}
\end{figure}

\begin{figure}
\resizebox{8cm}{!}{\includegraphics{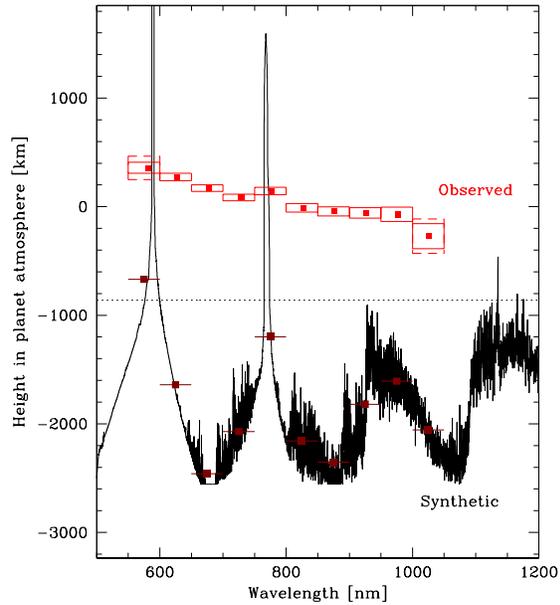}}
\caption{Atmospheric transmission spectrum of HD 189733b in the 600--1000 nm range. The vertical axis shows the height in the atmosphere of the effective transit radius of the planet, relative to an arbitrary reference level. The points are located at the intensity-averaged center of the 50-nm bins in wavelength, for our reference solution. The boxes  show the extent of the wavelength bins horizontally and the photon noise vertically. Dotted boxes correspond to the level of instrument systematics when larger than the photon noise. The solid line shows the synthetic spectrum of \citet{tin07}, normalised on the mid-infrared Spitzer data, with dots marking averages over 50-nm bins. The dotted line shows the infrared radius discussed in the text. }
\label{figspec}
\end{figure}

The time series in each 50-nm chromatic passband from 550 nm to 1050 nm are shown in Fig.~\ref{figchrom}. The transit depth was multiplied by 0.993 (first visit) and 0.998 (second visit), to account for the effect of unocculted spots, the sequence of the first transit affected by the planet occulting a large spot is not included and the second transit is corrected for the effect of the small spots (see Section~\ref{spots} on the effect of star spots). 

\begin{table}
\centering
\begin{tabular}{c r}
Passband [nm] &    Transit radius ratio   \\ \hline \hline
550 -- 600 &   0.156903 $\pm$   0.000095\\
600 -- 650 & 0.156744 $\pm$      0.000065\\
650 -- 700 &  0.156552 $\pm$     0.000057\\
700 -- 750 & 0.156388 $\pm$      0.000059\\
750 -- 800 & 0.156501 $\pm$      0.000064\\
800 -- 850 & 0.156210 $\pm$      0.000073\\
850 -- 900 & 0.156147 $\pm$      0.000081\\
900 -- 950 & 0.156120 $\pm$      0.000092\\
950 -- 1000 & 0.156097 $\pm$      0.000125\\
1000 -- 1050 &  0.155716 $\pm$      0.000218\\ \hline
\end{tabular}
\caption{Transmission spectrum of HD 189733b -- transit radius ratio in ten wavelength passbands, from our reference solution with correction for unocculted spots, with photon-noise errors. Systematic uncertainties are discussed in the text.}
\label{tablespec}
\end{table}

Table~\ref{tablespec} gives our `reference' transmission spectrum for HD 189733b, in terms of transit radius ratio ($R_{\rm pl}/R_{\rm star}$) in ten 50-nm passbands. 
Figure~\ref{figspec} displays the results in terms of height in the atmosphere at which the planet becomes opaque, relative to a fiducial height of 82000~km (using $R_{\rm star}=0.755 R_\odot$ from Paper~1). The boxes indicate, vertically, the photon-noise error bar, and horizontally the width of the passbands. The points are not centered on the boxes because the intensity-weighted wavelength means are different from the arithmetic means. Dotted boxes show the amplitude of the telescope/detector systematics, when larger than the photon noise. Note that the scale height of the atmopshere of HD189733b is of the order of 150 km. The uncertainties on the transit radius in 50 nm bands correspond to $\sim$ 50 km near the central wavelengths.

The cloud-free planetary atmosphere model from \citet{tin07}, normalized on the Spizer infrared data, is plotted for comparison. Strong absorption features from sodium (589 nm Na I doublet), potassium (769 nm K I doublet) and water are predicted. The horizontal dotted line shows the transit radius value of \citet{knu07} at 8 microns, which also corresponds to the highest level of the NICMOS transit radius measurements near 2 microns (M. Swain, private communication).

The main overall property of the 0.55 - 1.05 micron transmission spectrum of the planet HD189733 from our data is the lack of large features, such as would be expected from sodium, potassium and water in a transparent hot H$_2$ atmosphere. The spectrum is relatively flat at the average level of $R_{\rm pl}/R_*=0.1564$. 

Two more detailed observed properties are noted: a smooth slope of $\sim$1 \% in transit radius from 600 to 1000 nm, and a bump in the 750-800 nm passband. These two features can be affected by systematics. Nevertheless, they are robust in the face of our tests on these systematics. The value of the slope, in kilometers of transit radius, is 472 $\pm $ 42 (random) $\pm$ 106 (syst) kilometers over a 400 nm interval. Flattening the slope of the spectrum could be done by invoking unocculted spots, but this would require spot levels that would leave signatures that are not observed in the variability pattern and transmission spectrum.

The mean transit radius in the 600-1000 nm range is more than 2000 km higher than predicted by the cloud-free models of \citet{tin07}, and about 1000 km higher than the transit radius in the mid infrared. The uncertainty on the mean level is $\pm$ 16 (random) $\pm$ 61 (syst) km. The systematic errors are estimated by the quadratic addition of all sources of systematics studied in Section~\ref{ana}.

Slope changes at the 1-$\sigma$ level are observed at both edges of the spectra, and correspond to the expected position of the sodium line on the blue side, and the edge of a water band on the red side. Detailed examination of the pixel-by-pixel spectrum shows no indication that these features are due to narrow absorption features. They may be caused by the stronger systematics present at both edges.

No significant signatures of the sodium and potassium lines are detected. To quantify the absence of these lines, we repeated the extraction in 15-nm bands centered on the sodium line at 589 nm and the potassium line at 769 nm, compared to a baseline filter of 75 nm around the line positions. We measure a flux excess of $(-6.9 \pm 7.9) \times 10^{-5}$ in the sodium band (extra absorption during transit would yield a positive value here), and $(1.6 \pm 5.2) \times 10^{-5}$ in the potassium band. The quoted errors take only the photon noise into account. Systematic uncertainties estimated from the actual pixel-to-pixel scatter of the spectrum gives $\pm 11.9 \times 10^{-5}$ and $\pm 21.9 \times 10^{-5}$ for the sodium and potassium bands respectively.
Note that the 4-sigma detection of sodium for HD~209458 \citep{cha00} was with a narrow 1.2 nm band, their result with a 10 nm band was an insignificant $(3.1 \pm 3.6) \times 10^{-5}$ dimming during transit. In the case of HD~189733, additional unocculted starspots would increase the effect of the sodium line in the apparent transmission spectrum, so that the observed absence of feature is an upper limit. The narrow-passband data show no indication that the bump in the wide-band spectrum is located at the position of the potassium line.  However, the wide-band feature is visible independently in the spectra from all three visits in the HST data. It could therefore be real.

The data clearly exclude lines with the strength predicted by the condensate-free models of \citet{tin07}. This, together with the lower transit radius in the infrared and the redward slope in the spectrum, strongly suggests absorption of the grazing stellar flux by a haze of condensates high in the atmosphere of HD~189733b.

\section{Discussion}

Dust and condensates strongly affect the emission spectrum of brown dwarfs \citep[e.g.][]{all01,mar02} with atmospheres similar in composition and temperature to close-in gas giant planets. Condensates of iron, silicates and Al$_2$0$_3$ are thought to be abundant in brown dwarf and hot Jupiter atmospheres \citep{for05}. With the slant viewing geometry of transit transmission spectroscopy, less abundant compounds could also become important.  \citet{for05} mentions Cr, MnS, Na$_2$ , Zn,  C, KCl and NH$_4$H$_2$PO$_4$ with condensation temperatures in the relevant range. Grains formed by such condensates are usually sub-micron in size \citep{hel06}, producing a continuum absorption peaking in the visible and decreasing steeply in the near infrared.  In contrast to brown dwarf atmospheres, hot planet atmospheres are heated from above by the irradiation of their host star, so that non-equilibrium photochemical products could also play a role \citep{lia04}, like the ozone layer on Earth.

The observation of an almost featureless spectrum between 550 and 1050 nm for the planet HD 189733b is suggestive of absorption by condensates high in the atmosphere. The two strong alkali atomic absorption features expected in hot Jupiter atmospheres are either absent or much weaker than predicted, which can also be explained by absorption of the grazing stellar light by clouds or hazes\footnote{In this context we make no fundamental distinction between clouds and haze, which differ only by their opacity.} before the atomic lines can become efficient. The same explanation has been proposed to explain the weak strength of the sodium line in HD 209458 \citep{cha00,sea00}. 

The ACS, NICMOS and Spitzer data indicate that the transit radius in the near infrared is about 1000 km lower than in the visible. If confirmed, this difference would be consistent with the presence of condensates. Particles efficiently scatter radiation with wavelengths comparable to their radius, then the cross-section drops rapidly. The redward slope of the 550-1050 nm transmission spectrum suggests that the particles in the haze have sub-micron scales, so that their cross-section will be much lower in the near infrared. Atmospheres opaque at visible wavelengths but much more transparent in the near infrared are known in the Solar System. On Venus and Titan for instance, condensates block the visible light very high in the atmosphere, but much deeper layers, or even the planetary surface itself, can be seen in some spectral windows beyond 2 microns. Industrial hazes on Earth also share similar properties.


When relating the present results to the reflection and emission spectrum of HD 189733b measured during the secondary eclipse, several caveats are necessary. First, the transmission spectrum is formed solely along the limb of the planet, so that the conditions probed are only those of the day-night terminator. Second, the convection on close-in gas giant planets is expected to operate on a global scale, because the synchronous rotation lowers the Coriolis forces compared to, for instance, Jupiter. It is therefore concievable that the transmission spectrum varies from transit to transit, according to ``weather'' in the upper atmosphere. And third, a low opacity sufficient to scatter grazing starlight may not represent a dominant contribution with perpendicularly incident light. 

With these caveats, the presence of condensates in the upper atmosphere could influence the reflection/emission spectrum observed during secondary eclipse, through the scattering of light as well as by changing the pressure-temperature profile. Since most of the flux of the host star is emitted in the red, absorption and scattering by the haze is likely to modify the temperature profile of the atmosphere by injecting heat at high altitudes.  In a similar context, a temperature inversion has recently been postulated in the atmosphere of the transiting planet HD 209458b by \citet{bur07} to explain the emission spectrum recently observed by \citet{knu08} during the secondary eclipse. 

As a concluding remark, we note that our results indicate that the sunset over HD 189733b is red.

\section*{Acknowledgements}

The authors wish to thank Mark Swain and Giovanna Tinetti for very helpful discussions, and the anonymous referee for numerous suggestions that helped improve the manuscript.

\bibliography{artspec}

\end{document}